\documentclass[final,3p,times,twocolumn]{elsarticle}

\usepackage{graphics}
\usepackage{epsfig}
\usepackage{color}
\usepackage{amssymb}
\usepackage{lineno}
\usepackage{verbatim}

\journal{Advances in High Energy Physics}

\begin{document}

\begin{frontmatter}

\title{Analytic approximation of energy resolution in cascaded gaseous detectors}

\author[1]{Dezs\H{o} Varga}

\address[1]{Institute for Particle and Nuclear Physics, Wigner Research Centre for Physics of the HAS \\
29-33 Konkoly-Thege Mikl\'os Str., H-1121 Budapest, Hungary}

\begin{abstract}
An approximate formula has been derived for gain fluctuations in cascaded gaseous detectors such as GEM-s, based on the assumption that the charge collection, avalanche formation and extraction steps are independent cascaded processes. In order to test the approximation experimentally, a setup involving a standard GEM layer has been constructed to measure the energy resolution for 5.9 keV gamma particles. The formula reasonably traces both the charge collection as well as the extraction process dependence of the energy resolution. Such analytic approximation for gain fluctuations can be applied to multi-GEM detectors where it aids the interpretation of measurements as well as simulations.
\end{abstract}

\begin{keyword}
gaseous detectors \sep energy resolution \sep gain fluctuation

\end{keyword}

\end{frontmatter}


\section{Introduction}
\label{sec:Introduction}

The energy resolution of a gaseous detectors, that is, the precision with which a specific energy deposit can be measured, is a fundamental parameter and indication of instrumental quality. From the broad literature of gaseous detectors, such as classical \cite{bib:blumrolandi} or recent review documents \cite{bib:peskov}, one can understand the key contributing factors to the energy resolution. The fluctuation from the ionization process defines the ``intrinsic'' limit, quantified by the ``Fano factor''. The main contribution however stems from the fluctuation of the amplification of single electrons: as avalanches are formed, the amplitude (detected signal) fluctuates. A typical energy resolution for an MWPC is around 20\% FWHM, or equivalently, 8.5\% RMS, for the 5.9keV gamma line of the Fe-55 isotope. 

The energy resolution formula includes the single avalanche fluctuation: the signal amplitude $A$, usually expressed in units of input electrons, will have the following relative precision:

\begin{equation}
\left( \frac{\sigma_A}{A} \right)^2= (f^2 + (Fano))/n_{primary} + \sigma_{instrumental}^2
\label{eq:resolution}
\end{equation}

Here the key is $f$, the “single electron avalanche fluctuation”, defined as the standard deviation relative to the mean value. $f$ depends not only on gas composition but amplification geometry. The energy resolution scales with the primary electron number $n_{primary}$ from the ionization, whereas instrumental effects further broaden the response.

Assuming that an avalanche process (which can be quite complicated) initiated by a single electron finally results in N detected electrons with probability of $p(N)$, one can express $f$. The mean gain is the mean value $\bar{N}$ of the probability distribution. If the standard deviation is $\sigma_N^2 = {\bar{N^2}} - {\bar{N}^2}$, the relative fluctuation $f$ is defined as $f=\sigma_N / \bar{N}$. As an example, if the electron number distribution $p(N)$ is purely exponential, then $f=1$. 

Experimental and theoretical estimates for $f$ has been a subject for many studies. For proportional chambers, Alkhazov predicts 0.8-0.1 \cite{bib:alkhazov}, depending on gain. Micro-pattern Gaseous Detectors (MPGD-s) perform considerably better in energy resolution, due to the smaller $f$ values: this is true for GEM-s \cite{bib:schindlerthesis}, whereas Micromesh detectors are particularly outstanding \cite{bib:mesh} in this respect. 

For GEM detectors involving multiple amplification layers, the $f$ is referring to the whole process of the cascaded amplification. The electrons undergo various steps until the avalanche reaches the readout anode: they can be lost before entering the GEM holes, or may end up on the other side of the GEM, following the field lines. The question addressed in this paper: can we approximate $f$ for a cascaded process? How to include collection and extraction (in)efficiencies? This question is not only related to detector physics, but has relevance for practical instrument design, where simulations and measurements have a very broad range of parameters. Direct calculation can help in finding guidelines and assist search for optimized designs. We will find that the approach can be applied to any cascaded gaseous detectors, including hybrids of different MPGD types.

\section{Collection and extraction from a GEM layer}
\label{sec:collectionandextraction}

\begin{figure}[htb!]
        \centering
        \includegraphics[width=0.35\textwidth]{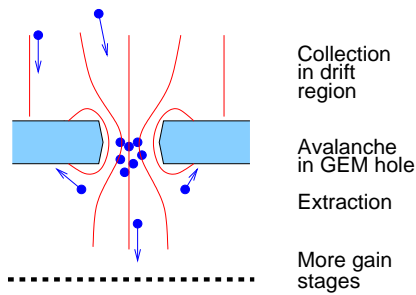}
        \caption{Cartoon for avalanche amplification processes in GEM. Blue dots represent electrons, which follow the field lines}
        \label{fig:gemcartoon}
        \end{figure}

The avalanche process in a GEM detector takes place in the high field of the GEM holes. The focusing field geometry guides the electrons to the holes, whereas the field lines which emerge from the hole extracts a certain fraction of the electrons \cite{bib:gem}. The situation may be show in the cartoon of Figure \ref{fig:gemcartoon}. In this sense the avalanche formation can be separated to four steps:

(1) Collection: the (single) electron from an ionization process drifts towards the GEM, being focused to a hole after leaving the approximately homogeneous drift region with field strength of $E_{drift}$. As some field lines end up on the GEM, the probability of entering the hole is $c$. This means that $(1-c)$ fraction of the electrons are “lost” for the subsequent amplification.

(2) Avalanche: those electrons which arrive in the high field region undergo a Townsend-avalanche. The total number of electrons is $N$, with relative fluctuation of $f$.

(3) A fraction of electrons is “extracted”: with a probability of $t$, depending strongly on the transfer field below the GEM, the electron leaves the vicinity of the hole and drifts towards the subsequent gain stages. Actually most of the electrons, $(1-t)$ fraction, end up on the bottom of the GEM.

(4) The extracted electrons will be further amplified leading to signal formation in the detector. For this step, the relative fluctuation will be denoted by $F$. 

The “effective” gain of the GEM, that is, the ratio of the electrons extracted relative to the number of those which arrived from the drift region, is G=cNt. Evidently the same notation may be used for GEM-s which are in the intermediate steps of a cascade, and one can even use this formulation for a broad range of gaseous detectors.

The question arises: can we determine the gain fluctuation $f_{all}$ for the complete process, (1) - (4), knowing the two probabilities $c$ and $t$, the “intrinsic” gain fluctuation f and the fluctuation from the subsequent stages $F$?

\section{Fluctuation of a cascaded processes}
\label{sec:fluctuationcascaded}

\begin{figure}[htb!]
        \centering
        \includegraphics[width=0.5\textwidth]{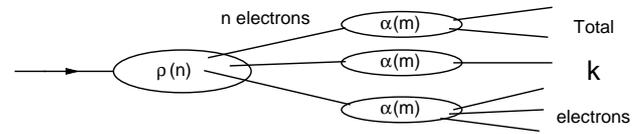}
        \caption{Cartoon for probability distributions in cascaded processes}
        \label{fig:statisticscartoon}
        \end{figure}

For cascaded process, one can calculate the fluctuation directly, derived from the probability distributions of each steps. This section gives a reasonable mathematical formulation which is useful to understand the specific result. Let us assume two cascaded steps. First, from a single electron, one creates $n$ electrons with probabilities of $\rho(n)$. From each of these electrons, the second step creates $m$ electrons with probabilities of $\alpha(m)$. The “cascading” leads to $k$ electrons, as depicted in Figure \ref{fig:statisticscartoon}. The mathematical question is, how to calculate fluctuation of the combination, from the fluctuations of each steps $\rho$ and $\alpha$? 

Let us denote the probability distribution of the combined process as $p(k)$. The fluctuation (RMS/mean, as already introduced) from the second process is $f_{\alpha} = \sqrt{\bar{\alpha^2}-\bar{\alpha}^2}/\bar{\alpha}=\sigma_\alpha / \bar{\alpha}$, where $\bar{\alpha}$ and $\bar{\alpha^2}$ denote the first and second moments; a similar notation will be used for the first and combined processes, $\rho$ and $p$, resulting in their respective fluctuations of $f_{\rho}$ and $f_p$. 

In case that two electrons are emitted from the first process, the number of electrons after the second process is the convolution of the respective probability distribution, $\alpha(m)$, with itself:

\begin{equation}
(\alpha * \alpha) (k) = \sum_{m=0}^k \alpha(m) \alpha(k-m)
\end{equation}

whereas if $l$ electrons are emitted from the first process, the $l$-fold convolution of the $\alpha(m)$ distribution, denoted by $(\alpha)^{l*}(k)$ is to be taken. This latter contains multiple sums, and most easily experessed as an iterative formula

\begin{equation}
(\alpha)^{(l+1)*} (k) = \sum_{m=0}^k (\alpha^{l*})(m) \alpha(k-m)
\end{equation}

In the general case, the first process emits $n$ electrons with probability distribution of $\rho(n)$, therefore $p(k)$ can be directly calculated as follows:

\begin{equation}
p(k) = \sum_{n=0}^\infty \rho(n) (\alpha^{n*})(k)
\end{equation}

Direct evaluation of the fluctuation of $p(k)$, that is, $f_p=(\bar{p^2}-\bar{p}^2)/\bar{p}$ gives the key equation for the results in the present paper:

\begin{equation}
f^2_p = f^2_\alpha + \frac{1}{\bar{\alpha}} f^2_\rho
\label{eq:cascfluc}
\end{equation}

where $\bar{\alpha}$ is the mean number of electrons from the first process. For completeness, one can also derive an “expected” result, stating that $\bar{p} = \bar{\alpha} \bar{\rho}$: that is, the mean number of electrons from the cascaded process is the product of the mean electrons from the individual steps.

Equation \ref{eq:cascfluc} states that the cascaded process inherits its fluctuation from the first step, and increased with adding in quadrature the fluctuation of the second step, suppressed by the “gain” of the first step. The formula resembles, but distinct from, the summing of independent variables. Equation \ref{eq:cascfluc}, which is conveniently simple, can be used to evaluate the cascaded gain fluctuation.

\section{Fluctuation from specific processes in GEM-s}
\label{sec:processes}

The collection, avalanche and extraction steps can be seen formally as a “cascaded” statistical process, therefore the formulation introduced can be applied. One needs however to directly calculate the corresponding fluctuations.

\subsection{Collection}
\label{sec:collection}

Starting out from a single electron, the probability of capturing or loosing the electron is $c$ and $(1-c)$ respectively. This means that the probability distribution is

$$p_{coll}(n)=\left\{\begin{array}{cc}1-c & \textrm{for} \, \, \, \, n=0 \\ c & \textrm{for} \, \, \, \, n=1 \\ 0 & \textrm{otherwise} \end{array}\right\}$$

The mean is certainly $c$, whereas the fluctuation $f_{coll}^2 = 1/c-1$. 

\subsection{Avalanche}
\label{sec:avalanche}

The mean number of electrons created in the avalanche, as introduced earlier, is $N$, whereas the fluctuation of the avalanche is denoted by $f$. Both values we take as parameters.

\subsection{Extraction}
\label{sec:extraction}

The fluctuation from the extraction can be calculated similarly to the collection process. Assuming a single electron just leaving the avalanche region, it is extracted to the transfer region with probability t, that is:

$$p_{extr}(n)=\left\{\begin{array}{cc}1-t& \textrm{for} \, \, \, \, n=0 \\ t & \textrm{for} \, \, \, \, n=1 \\ 0 & \textrm{otherwise} \end{array}\right\}$$

Again, the mean of the probability distribution is $t$, whereas its fluctuation $f_{extr}^2=1/t-1$. 

\subsection{Formula for GEM-s}
\label{sec:gemformula}

Having the individual processes, one can set up the cascade (1) - (4) to calculate the total fluctuation. For collection (1), we have

$$(1/c-1) = f_{coll}^2$$

The next step of the cascade is the avalanche (2), which is to be combined, according to Equation \ref{eq:cascfluc}, in the following way

$$(1/c-1) + 1/c f^2 = f_{coll}^2 + 1/c f^2$$

Combining (1)+(2)+(3), one gets

$$(\frac{1}{c}-1) + \frac{1}{c} f^2 + \frac{1}{NC} (\frac{1}{t}-1)$$

This is actually a “single GEM contribution”. And finally the complete process (1)+(2)+(3)+(4), including the contribution from all subsequent stages:

$$f_{all}^2=(\frac{1}{c}-1) + \frac{1}{c} f^2 + \frac{1}{Nc} (\frac{1}{t}-1) + \frac{1}{Nct} F^2$$

Having introduced the effective gain $G=cNt$, one can rewrite the formula in the following convenient way:

\begin{equation}
f_{all}^2=(\frac{1+f^2}{c}-1) + \frac{1}{G} (1-t + F^2)
\label{eq:formula}
\end{equation}

In this formula, the first term of the sum contains collection, whereas the second term, being suppressed by $1/G$, contains extraction. The values of $f$ and $F$ are around 1. The surprisingly simple formula gives an easy access to experimental comparison.

\section{Measurements of energy resolution at 5.9 keV}
\label{sec:measurements}

\subsection{Experimental setup}
\label{sec:expsetup}

\begin{figure}[htb!]
        \centering
        \includegraphics[width=0.4\textwidth]{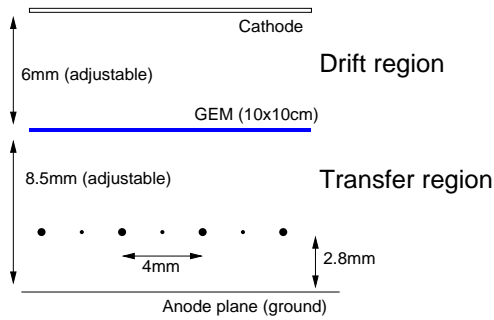}
        \caption{Detector outline, with a single GEM and an MWPC as a high gain stage}
        \label{fig:chamberoutline}
        \end{figure}

In order to obtain a systematic and consistent measurement set, a simple experimental setup has been constructed, including a single GEM layer, and an MWPC as a high gain stage. The setup is shown in Figure \ref{fig:chamberoutline}, indicating the detector geometry. The GEM is a “standard” CERN product, 10cm by 10cm in size, with 140 micron pitch, 70 micron hole diameter and 60 micron thickness. For the MWPC, a version of a ``Close Cathode Chamber'' \cite{bib:ccc} has been applied.

\begin{figure}[htb!]
        \centering
        \includegraphics[width=0.5\textwidth]{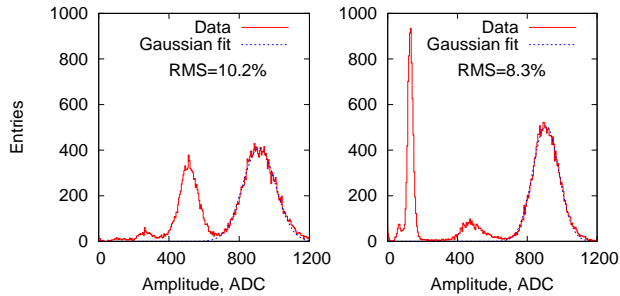}
        \caption{Example of a 5.9keV peak in the detector, in argon-based gas mixture.}
        \label{fig:argonexample}
        \end{figure}

\begin{figure}[htb!]
        \centering
        \includegraphics[width=0.5\textwidth]{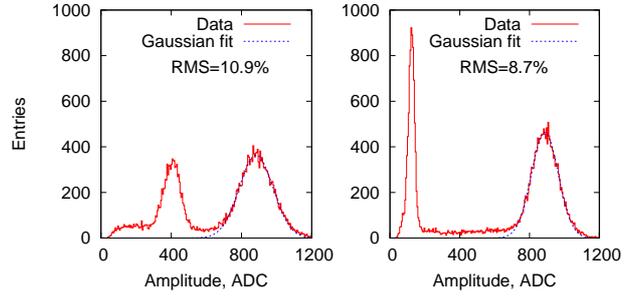}
        \caption{Example of a 5.9keV peak in the detector, in neon-based gas mixture}
        \label{fig:neonexample}
        \end{figure}

The measurements have been performed in two gas mixtures, $Ar+CO_2$ at 80:20 and $Ne+CO_2$ at 90:10 proportions. Using an Fe-55 isotope as X-ray source for 5.9 keV gamma particles, examples of amplitude spectra are shown in Figure \ref{fig:argonexample}. One can clearly observe the expected structures: the main 5.9keV peak, the escape peak in Ar, and also the peak corresponding to conversions below the GEM (that is, only amplified by the MWPC). Figure \ref{fig:argonexample} and \ref{fig:neonexample} shows Gaussian fits as examples, demonstrating that clear evaluation of the energy resolution is achievable with the presented setup.

\begin{figure}[htb!]
        \centering
        \includegraphics[width=0.4\textwidth]{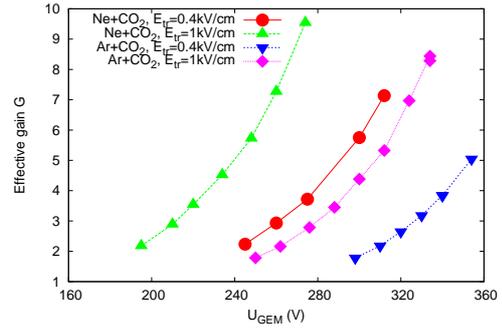}
        \caption{Effective gain G as a function of GEM voltages for the various measurements}
        \label{fig:effgainvsugem}
        \end{figure}

The GEM effective gain can be directly evaluated by taking the ratio of the peak position from gamma conversion above and below the GEM. The effective gain as a function of GEM voltage in the various gases is shown in Figure \ref{fig:effgainvsugem}. The measurements were taken at two different transfer fields, value of 0.4kV/cm and 1kV/cm.

\subsection{Resolution versus drift: access to collection process}
\label{sec:resvscoll}

\begin{figure}[htb!]
        \centering
        \includegraphics[width=0.5\textwidth]{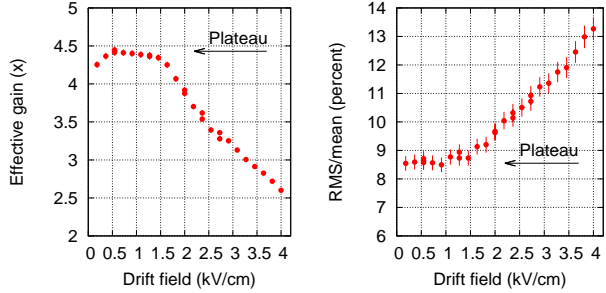}
        \caption{The effective gain (left) and the energy resolution (right) dependence on the drift field}
        \label{fig:cathodedep}
        \end{figure}

The collection efficiency depends on the drift field \cite{bib:gem}. The collection efficiency $c$ can be approximated by measuring the gain as a function of the transfer field, shown in Figure \ref{fig:cathodedep}. One clearly observes a plateau up to about 1.6kV/cm, which is a starting point of a gradual decrease: here $c$ can be estimated as the gain change relative to the plateau. The resolution, shown in the right panel of Figure \ref{fig:cathodedep}, features a corresponding structure: an increase (broadening) of the energy resolution.

\subsection{Resolution versus gain: access to the extraction process}
\label{sec:resvsextr}

\begin{figure}[htb!]
        \centering
        \includegraphics[width=0.5\textwidth]{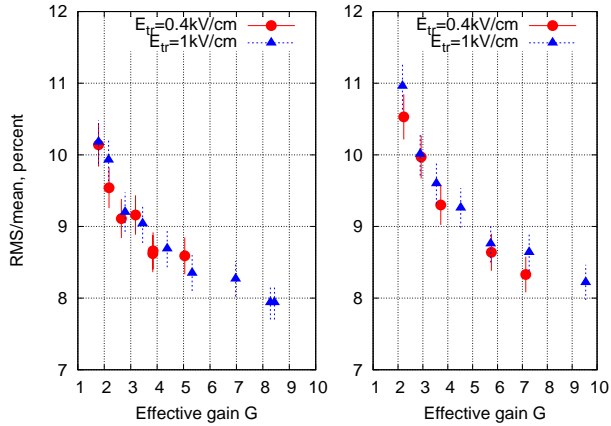}
        \caption{The measured energy resolution (RMS) as a function of effective gain. Note the improvement of the resolution with gain, both for the argon-based mixture on the left, and for the neon-based mixture on the right}
        \label{fig:sigmavseffgain}
        \end{figure}

The energy resolution of the 5.9keV peak has been evaluated for the two transfer field values, at fixed drift field of 0.8 kV/cm (that is, safely on the collection plateau with c=1), as a function of GEM effective gain. The results, shown in Figure \ref{fig:sigmavseffgain}, show a decreasing trend both for Ar and Ne. It is indicative to observe that the resolution is nearly the same for the different transfer fields: indeed from Equation \ref{eq:formula}, one expects that the effective gain, $G$, is the most relevant parameter in the second term of the formula.

\section{Predictive power of the analytic formula}
\label{sec:predictivepower}

In this section, the quantitative comparison of the energy resolution will be given between that predicted by Equation \ref{eq:formula} and the measurements. Since this paper is concerned more about the formulation and trends, rather than the specific parameters, some assumptions are needed. First, for Equation \ref{eq:resolution}, the Fano factor will be set to 0.2 \cite{bib:peskov}, whereas the primary electron number for Ar+CO2 and Ne+CO2 mixture is assumed to be 210 and 165 respectively \cite{bib:peskov}. The $F$, that is, the gain fluctuation in the MWPC stage is assumed to be $F=1.0$ (pure exponential). The extraction efficiency $t$ is well below 1, and various values between 0 and 0.2 will be assumed, values which are actually reasonable \cite{bib:gem} for the experimental transfer fields. An important, and actually the most interesting figure from the point of view of detector physics is $f$, the “intrinsic” avalanche fluctuation of GEM-s. In this paper, we will tentatively use 0.9 for Ar+CO2, and 0.8 for Ne+CO2, figures which are motivated by simulations \cite{bib:schindlerthesis}, and will be subject of our own detailed later studies as well.

The energy resolution calculation requires the instrumental fluctuation $\sigma_{instrumental}$ ; this is now set to 2\%, which is estimated from the electronics and digitization noise, and is therefore a lower limit. We will observe however, that the various parameters are strongly correlated as seen in the formula, and most of the qualitative predictions will not depend on the specific parameter settings.

\subsection{Energy resolution decrease with collection loss}
\label{sec:resdecrvscoll}

\begin{figure}[htb!]
        \centering
        \includegraphics[width=0.4\textwidth]{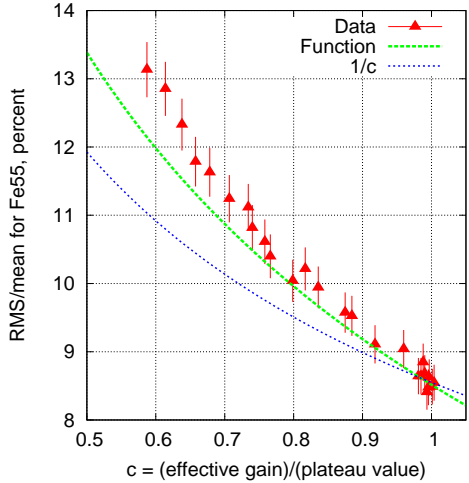}
        \caption{The measured energy resolution at 5.9keV, in $Ar+CO_2$, as a function of the estimated collection efficiency. The curves show the prediction by Equation \ref{eq:formula}, as well as a proportionality to $\sqrt{1/c}$}
        \label{fig:cathodepred}
        \end{figure}

The most prominent feature of the resolution broadening pattern is the correlation with loss of collection. From Figure \ref{fig:cathodedep} one can estimate the collection fraction c by dividing the actual gain with the plateau gain. Using this value, the measured energy resolution as a function of c is shown in Figure \ref{fig:cathodepred}. Comparing this to the prediction from Equation \ref{eq:formula}, shown as the estimated function, a reasonable agreement is apparent. The “intuitive” guess would be proportionality to $\sqrt{1/c}$ (if half of the charge is collected, the fluctuation increases by $\sqrt{2}$ ), whereas the measurements are considerably higher. One can note that the data is actually higher than the prediction: this may be due to a “correlated” loss of the charge cluster, that is, if the assumption that the avalanche processes are independent (leading to Equation \ref{eq:cascfluc}) is not exactly correct. In conclusion, the predicted decrease of the resolution is well corresponding to the actual measurements.

\subsection{Energy resolution decrease with gain}
\label{sec:resdecvsgain}

\begin{figure}[htb!]
        \centering
        \includegraphics[width=0.5\textwidth]{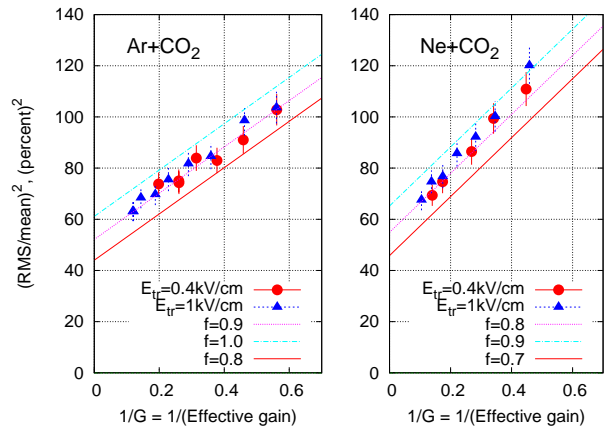}
        \caption{The measured energy resolution (RMS) as a function of $1/G$. }
        \label{fig:sigmavseffgainpred}
        \end{figure}

The formula, in case of $c=1$ (full collection), predicts a dependence on $G$, with other parameters being not so relevant individually (such as $t$, or increase of $f$ with GEM voltage). According to Equation \ref{eq:formula}, one should get a linear dependence of the square of the width on $1/G$. The corresponding measurement, along with the prediction from Equation \ref{eq:formula} and Equation \ref{eq:resolution} is shown in Figure \ref{fig:sigmavseffgainpred}.

Both for Ne and Ar based gases, the prediction gives a correct slope against $1/G$, and indicates that the gain indeed follows the rule from Equation \ref{eq:formula}. The Figure shows the prediction for various values of $f$, which means that the linear behaviour, and the steepness of the slope, is largely independent of the specific $f$ value.

\section{Conclusions}

Interpreting the avalanche formation in GEM-s as a statistical process, one can calculate the gain fluctuation from the fluctuations of individual steps. This includes charge collection, avalanche development, extraction and subsequent amplification. The final fluctuation, directly related to energy resolution, can be expressed as a formula represented by Equation \ref{eq:formula}. Experimentally two steps, the collection and the extraction, have been studied systematically. Estimating the collection efficiency from the gain loss at varying drift fields, the energy resolution can be reasonably predicted for 5.9keV gamma rays. Similarly, in case of varying extraction field, a specific dependence of the resolution on $1/G$ is predicted, which is supported by the measurements. The approximate formula for energy resolution can be used as a basis for estimating merits of figure for gaseous detectors, especially when energy resolution worsening will balance improvement in other parameters such as ion blocking or overall gain. The formulation also helps to understand the avalanche formation, as a statistical process in cascaded gaseous detectors, and thus simplifies the interpretation of experimental or simulated data.

\section{Acknowledgements}

This work has been supported by the “Momentum” grant of the Hungarian Academy of Sciences (LP2013-60). I wish to thank the support from the REGARD group members at the Wigner Research Centre in Budapest.

\bibliographystyle{model1a-num-names}

\end{document}